\documentclass[twocolumn,letterpaper,prb]{revtex4}
\usepackage{amsmath,amsthm,amscd}
\usepackage{dcolumn}
\usepackage{graphicx}
\usepackage{epsfig}

\begin{document}
\title{Theory of cavity-polariton self-trapping and optical strain in polymer chains}
\author{M. V. Katkov, Y. V. Pershin and C. Piermarocchi}
\affiliation{Department of Physics and Astronomy, Michigan State
University, East Lansing, Michigan 48824}

\begin{abstract}
We consider a semiconductor polymer chain coupled to a single electromagnetic
mode in a cavity. The excitations of the chain have a mixed exciton-photon
character and are described as polaritons. Polaritons are coupled to the
lattice by the deformation potential interaction and can propagate in the
chain. We find that the presence of optical excitation in the polymer induces
strain on the lattice. We use a BCS variational wavefunction to calculate the
chemical potential of the polaritons as a function of their density.  We
analyze first the case of a short chain with only two unit cells in order to
check the validity of our variational approach. In the case of a long chain and
for a strong coupling with the lattice, the system undergoes a phase transition
corresponding to the self-trapping of polaritons. The role of the exciton
spontaneous emission and cavity damping are discussed in the case of
homogeneous optical lattice strain.
\end{abstract}

\date{\today}

\maketitle

\section{Introduction}
Recent advances in nano-spectroscopy have demonstrated the possibility of
addressing a single conjugated polymer chain in a solid
matrix.~\cite{muller03,guillet01} These optical studies of single chains
overcome limitations due to ensemble averaging and inhomogeneous broadening and
give a clear picture of the exciton dynamics. Most interestingly, a 1D
singularity in the optical density of states, a $T^{1/2}$ temperature
dependence of the the exciton lifetime,~\cite{dubin02} and a macroscopic
quantum spatial coherence~\cite{dubin06} have been observed in isolated red
polydiacetylene chains. These features suggest that the behavior of excitons in
polymer chains can be very close to that of an ideal semiconductor quantum wire
system.

In this paper, we study excitons in a polymer chain coupled to a
single electromagnetic mode in a cavity. The resulting mixed states
of excitons and cavity photons can be describes in terms of
polariton quasiparticles. We will focus on the properties of
polaritons in the presence of a deformation potential interaction
with the lattice. We will consider a modified Su-Schrieffer-Heeger
(SSH)~\cite{heeger88} model, describing the propagation of excitons
in the polymer~\cite{wilson83}, with an additional term that takes
into account the coupling with a single cavity mode. Excitons are
modeled as excitation of two level systems localized within the unit
cells of the polymer chain, and a variational mean field approach is
used to calculate the properties of the ground state of the systems
at zero temperature as a function of the density of polaritons. A
similar approach was used to study the transition between a
polariton Bose Einstein Condensate (BEC) to a laser-like behavior
for polaritons in a cavity.~\cite{szymanska02} The coupling to the
lattice by the deformation potential adds new features to the ground
state properties of polaritons. We will show that a
self-trapping~\cite{toyozawa03} of polaritons occurs at a threshold
value for the polariton density. This mechanism could give rise to a
BEC of polaritons which localize spontaneously without the need of
external traps or strain fields. Compared to the semiclassical
treatment we have used in the case without cavity,~\cite{katkov06}
the description in terms of cavity-polaritons give a clearer picture
of the physics of the self-trapping process, and introduces new
features related to the strong exciton-cavity photon coupling.
Although our model contains a single cavity mode, we expect the
result to be relevant also in the description of organic systems in
planar cavity geometries, for which interesting interplays of the
exciton-cavity and exciton- LO phonon dynamics have been
predicted.~\cite{agranovich03} Exciton-polaritons are one of the
most promising candidates for the realization of BECs in condensed
matter systems.~\cite{deng02} Exciton-polaritons
lasing~\cite{deng03} and matter-based parametric
amplifiers~\cite{saba01} are additional important applications of
these quasiparticles. Exciton-polaritons in organic systems are
particularly interesting due to the big excitonic oscillator
strength and to their strong coupling to phonons, which gives rise
to strong optical nonlinearities.~\cite{greene90} Evidence of
polaritonic effects in a single polydiacetylene chains has been
recently reported.~\cite{dubin06prb} We will focus on semiconductor
polymer chains with a non-degenerate ground state. A well known
example in this class of materials is
polydiacetylene~\cite{bloor85}. However, we will keep our theory
general in such a way that it can be extended to polymers with
similar properties.

The paper is organized as follows: in Sect.~\ref{sec:model} we
introduce the model. In Sec.~\ref{sec:energy} we illustrate the
variational approach used to calculate the ground state properties.
In Sec.~\ref{sec:twosites} we compare the variational results to an
exact calculation with two sites in order to establish the validity
of our approach. Sect.~\ref{sec:num} provides the details of the
numerical energy minimization procedure used to find the
distribution of polaritons in the chain and the total energy of the
systems. The results on the self-trapping phase transition as well
as analytical results that can be obtained is some limits are
described in Sec.~\ref{sec:chem}. Sec.~\ref{sec:bistab} analyzes the
role of the excitonic spontaneous emission and of the finite
Q-factor of the cavity. Conclusions are in
Sec.~\ref{sec:conclusions}.

\section{The model}
\label{sec:model} The model consists of the 1D system of excitons coupled to
lattice deformations as well as to a single cavity mode of the electromagnetic
field. The Hamiltonian can be written as
\begin{equation}
H=H_{D}+H_{SSH}~.\label{eq:htot}
\end{equation}
The first term corresponds to the Dicke model~\cite{dicke54} of an
ensemble of two level systems coupled to a single electromagnetic
mode
\begin{equation}
H_{D}=\omega_c c^{\dagger} c+\sum_n \frac{g}{2}(B^\dagger_n c+c^{\dagger} B_n)
+\omega_X\sum_n B^{\dagger}_n B_n~, \label{eqdicke}
\end{equation}
where $\omega_X$ is the exciton energy. We will use $\hbar=1$ throughout the
paper. $B^{\dagger}_n$, $B_n$ are operators of creation, annihilation of
excitons in a singlet spin state and $c^{\dagger}$, $c$ are creation,
annihilation operators for the cavity photons of energy $\omega_c$. Each site
$n$ of the model represents a single monomer of the polymer chain. The
parameter $g$ indicates the exciton-cavity dipolar coupling constant. In the
Dicke model, the atoms do not have translational degrees of freedom and there
is no direct transfer of optical excitation from one atom to the other. Here,
we need to add two additional features to the Dicke model: ({\it i}) the
excitations can hop from site to site and move along the backbone of the
polymer chain, and, ({\it ii}), the hopping of the excitation depends of the
relative position of the sites in the lattice, which can move around their
equilibrium position. This can be represented by an the excitation transfer of
the SSH form as
\begin{eqnarray}
H_{SSH}&=&\sum_{n}\frac{p_n^2}{2M} + \sum_n
\frac{C}{2}(u_{n+1}-u_n)^2 \nonumber \\
& & -\sum_n t_{n+1,n}(B^\dagger_{n+1}B_n +B^\dagger_n B_{n+1})~. \label{eqssh}
\end{eqnarray}
The SSH model was used to describe the electronic transport in
polyacetylene chains~\cite{heeger88}, and was extended to the case
of excitonic transport in polydiacetylene~\cite{wilson83}. In
Eq.~(\ref{eqssh}), $M$, $C$, $u_n$ and $p_n$ are the mass, elastic
constant, total displacement and momentum of the $n$th site of the
chain. The hopping term is $t_{n+1,n}=t_0-\gamma(u_{n+1}-u_n)$,
where $\gamma$ is related to the exciton-phonon deformation
potential $D=2 \gamma a$. $a$ is the site separation in the
tight-binding chain. The value of $t_0$ is determined by the exciton
effective mass $m$ as $ t_0=\frac{1}{2m a^2}$.

\section{Energy minimization}
\label{sec:energy} We consider a polariton trial wave function  which is a
product of a coherent state for photons and a BCS state for excitons as
~\cite{eastham01}
\begin{equation} |\lambda,\alpha, \beta,\varphi
\rangle=|\lambda\rangle \prod_{n}
 \left(\alpha_n|0\rangle_{n}+e^{i\varphi_n}\beta_n|1\rangle_{n}\right)~,
\label{var}
\end{equation}
where $\lambda$, $\alpha_n$, $\beta_n$, and $\varphi_n$ are variational
parameters, and $|\lambda\rangle$ represents a coherent state for the cavity.
The coefficients $\alpha_n$ and $\beta_n$ are subject to the single-occupancy
constraint
\begin{equation}
|\alpha_n|^2+|\beta_n|^2=1~.
\end{equation}
$|0\rangle$ is the vacuum state of the cavity mode, and
$|0\rangle_n$ and $|1\rangle_n$ denote the ground and the excited
state of the two level system at the site $n$. We assume $\lambda$
real and we fix the phase $\varphi_n$ to make $\alpha_n$ and
$\beta_n$ real. Due to the hopping, the variational coefficients
will depend on the index $n$ in the general case.

We can interpolate continuously the wavefunction by transforming the discrete
sum $\sum_{n=1}^{N}$ in a continuous integral $\int_0^N d\nu$. In this way, we
can define the optical polarization $ \psi(\nu)=2\alpha (\nu)\beta(\nu)$, where
$\alpha$ and $\beta$ are continuous functions of $\nu$. We can express the
total energy of the system as the expectation value of the Hamiltonian in
Eq.~(\ref{eq:htot}) with the trial wavefunction in Eq.~(\ref{var}). In the
continuum form, the total energy can be expressed as
\begin{widetext}
\begin{eqnarray}
 E&=&\omega_c\lambda^2+\frac{1}{2}\int_0^N \left[ \frac{p^2}{M}+
 C u'^2+t_0(\psi'^2+\psi^2 \varphi'^2)-t_0\psi^2 +\gamma u'\psi^2
 +g\lambda \psi \cos\varphi-\omega_X\sqrt{1-\psi^2}\right]d\nu~, \label{energy}
\label{eq:cont}
\end{eqnarray}
\end{widetext}
where $\varphi(\nu)$, $p(\nu)$ and $u(\nu)$ are also continuous
functions. The prime indicates the derivative with respect to the
variable $\nu$. In deriving Eq.~(\ref{eq:cont}) we have assumed that
$|\alpha(\nu)|>|\beta(\nu)|$ along the chain, which corresponds to
the condition of negative detuning between the exciton resonance and
the cavity mode. Notice that the total polariton number
\begin{equation}
N_{P}=\lambda^2 + \int|\beta|^2d\nu=\lambda^2+\frac{N}{2}-\frac{1}{2}\int
\sqrt{1-\psi^2}d\nu
\end{equation}
is a conserved quantity. In order to find the ground state of the
system at a fixed polariton density we perform a variational
minimization of $\langle H-\mu N_{P} \rangle$ with respect to the
functions $\psi$, $\varphi$, $u$ and with respect to the constant
$\lambda$. From the condition $\delta (E-\mu N_P)=0$ we obtain the
system of equations
\begin{subequations}
\label{eq:eu}
\begin{eqnarray}
t_0(\varphi''\psi^2+\varphi'{\psi^2}')+\frac{g \lambda}{2} \psi \sin\varphi =0~, \label{eq:eua}\\
Cu''+\frac{\gamma}{2}{\psi^2}'=0~, \label{eq:eub}\\
 2t_0\psi''+2t_0\psi(1-\varphi'^2)-2\gamma\psi u'- \nonumber\\
 -g\lambda\cos{\varphi}-\frac{(\omega_X-\mu)\psi}{\sqrt{1-\psi^2}}=0~, \label{eq:euc}\\
(\omega_c-\mu)\lambda+\frac{g}{4}\int\psi\cos{\varphi} d\nu=0~.
\label{lambdaeq}
\end{eqnarray}
\end{subequations}
From Eqs.~(\ref{eq:eua}) and~(\ref{eq:cont}) we can see immediately
that, in the assumption $\alpha \beta>0$, and $g>0$ we can take the
solution $\varphi=\pi$, since this variable has no constraint. That
makes the global phase of the optical polarization constant along
the chain and out of phase with respect to the cavity field. In
order to have a closed equation for $\psi$ we can integrate the
equation for $u$ in Eq.~(\ref{eq:eub}) and substitute in
Eq.~(\ref{eq:euc}). By direct integration of Eq.~(\ref{eq:eub}) we
can write
\begin{equation}
u'=-\frac{\gamma}{2C}|\psi|^2
+a\Delta~, \label{derofdisp}
\end{equation}
where $ \Delta$ is a dimensionless constant of integration. We
choose $\Delta=0$, which implies that the total length of the
polymer is not fixed. The force constant $C$ can be expressed in
terms of the sound velocity $S$ as $C=\frac{S^2M}{a^2}$.  Finally,
the system of equation in Eq.~(\ref{eq:eu}) can be rewritten in the
form
\begin{subequations}
\begin{eqnarray}
 -2t_0 \psi -t_0 \psi''-\chi\psi^3-g\lambda
 +\frac{(\omega_X-\mu)\psi}{\sqrt{1-\psi^2}}=0~,\label{eq:fina}\\
(\omega_c-\mu)\lambda-\frac{g}{4}\int\psi d\nu=0~,\label{eq:finb}
\end{eqnarray}
\label{final}
\end{subequations}
where the coefficient of the cubic term $\chi=\frac{D^2}{4MS^2}$.
This system of coupled equation will be solved numerically in
Sec.~\ref{sec:num}.

\section{Two sites model}
\label{sec:twosites} In order to check the validity of the BCS trial
wave function in Eq.~(\ref{var}) and the accuracy of the variational
approach, we compare in this section the exact solution of the
problem with a result obtained with the variational approach for two
lattice sites. The exact solution is obtained by writing the wave
function in the form
\begin{equation}
\Psi=\eta_0|00 N_P\rangle+\eta_1|01 N_P-1\rangle +\eta_2|10
N_P-1\rangle+ \eta_3|11 N_P-2\rangle~, \nonumber
\end{equation}
where the first term in the right-hand side corresponds to the state
with zero excitons and $N_P$ photons, the second and third terms
describe the states with one exciton and $N_P-1$ photons, and the
last term is a state with two excitons and $N_P-2$ photons. Here, we
assume that the number of polaritons in the system  $N_P>1$. In this
section we will also consider that the hopping $t_{12}$ as a fixed
parameter. Using this form for the wave function in the
Schr\"odinger equation with the Hamiltonian in Eq.~(\ref{eq:htot})
we obtain a set of equations for $\eta_i$:
\begin{eqnarray*}
i\dot \eta_0=\omega_c
N_P\eta_0+\frac{g}{2}\sqrt{N_P}(\eta_1+\eta_2)~,\;\; \label{alph0}
\\
i\dot \eta_1=\varepsilon_1\eta_1
+\frac{g}{2}(\sqrt{N_P}\eta_0+\sqrt{
N_P-1}\eta_3)-t_{12}\eta_2~,\;\;
\\
i\dot \eta_2=\varepsilon_1\eta_2 +\frac{g}{2}(\sqrt
{N_P}\eta_0+\sqrt{ N_P-1}\eta_3)-t_{12}\eta_1~,\;\;
\\
i\dot \eta_3=\left(\omega_c(
N_P-2)+2\omega_X\right)\eta_3+\frac{g}{2}\sqrt{N_P-1}(\eta_1+\eta_2)~,\;\;\label{alph3}
\end{eqnarray*}
where $\varepsilon_1=\omega_c(N_P-1)+\omega_X$. The energy spectrum
and, in particular, the ground state energy are found by direct
diagonalization.

Using the trial wave function approach described in the previous section, we find that
\begin{eqnarray}
\langle \alpha,\lambda,\beta,\varphi |H|
\alpha,\lambda,\beta,\varphi\rangle
=\omega_c\lambda^2+\omega_X(\beta_1^2+\beta_2^2)- \nonumber \\
-2t_{12}\alpha_1\beta_1\alpha_2\beta_2\cos(\varphi_1-\varphi_2)+g\lambda\sum\limits_{i=1}^2
\alpha_i\beta_i \cos\varphi_i~. \label{avham}
\end{eqnarray}
Without a loss of generality, the following substitutions are made
in Eq. (\ref{avham}): $\varphi_i=\pi$,
$\lambda=\sqrt{N_P-\beta_1^2-\beta_2^2}$,
$\alpha_i=\sqrt{1-\beta_i^2}$. The  resulting expression depends
only on $\beta_i$. The minimum value of $\langle
\alpha,\lambda,\beta,\varphi |H|
\alpha,\lambda,\beta,\varphi\rangle$ is considered as the ground
state energy.
\begin{table}
\begin{tabular}{c|c|c}
$N_P$& $E_0$, exact & $E_0$, trial WF \\ \hline 4 & 2.993 & 3.089
\\
 20 & 16.928 & 16.972 \\
 100 & 87.837 & 87.857 \\
 200 & 177.012 & 177.027 \\
 1000 & 893.528 & 893.521
\end{tabular}
\caption{Ground state energies calculated exactly and using the
trial wave functions. The calculations were made using the following
set of parameters: $\omega_c=0.9E_g$, $t_{12}=0.5E_g$, $g=0.2E_g$.}
\label{table1}
\end{table}
Table \ref{table1} shows the comparison between the exact and the
variational results. The ground state energy was calculated for
different values of the photon number in the cavity. The ground
state energies calculated with the variational approach are in good
agreement with the exact calculation. As expected, the exact ground
state energy is  slightly smaller than the energy calculated using
the trial wave function. This difference can be related to
correlation effects not included in the trial wave function. Also
notice that the difference between the ground state energy
calculated with the two approaches decreases at a larger number of
the photons. In fact, the correlation effects between the excitons
and the cavity photons are expected to disappear for a photon number
much larger that the excitation number in the system.

\section{Numerical procedure}
\label{sec:num} Once established the validity of the variational
approach, we can study the general case of a long chain. In order to
study on the same ground the inter-cell hopping and the presence of
the cavity field we solve numerically the system of
Eqs.~(\ref{final}). We use the steepest descent method of functional
minimization.~\cite{SCP} This method is efficient in solving
numerically Gross-Pitaevskii equations for BECs,~\cite{dalfovo96}
which are similar to our nonlinear equations. The method consists of
projecting an initial trial state onto the minimum of an effective
energy ${\cal H}$ by propagating the state in an imaginary time. We
start from an initial field parameter $\lambda(\tau=0)$ and a trial
function $\psi(\tau=0)$, then $\lambda(\tau) $and $\psi(\tau)$ are
evaluated in terms of the equations
\begin{equation}
\frac{\partial \psi(\tau)}{\partial \tau}=-\frac{\overline{\delta} {\cal
H}}{\overline \delta \psi_n(\tau)}\label{numpsi}
\end {equation}
and
\begin{equation}
\frac{\partial \lambda(\tau)}{\partial \tau}=-\frac{\overline{\delta} {\cal
H}}{\overline \delta \lambda(\tau)}~,\label{numlambda}
\end {equation}
where $\overline{\delta}$ indicates a constrained derivative that
preserves normalization. Eqs.~(\ref{numpsi}) and (\ref{numlambda})
define a trajectory in the parameter space for the optical
polarization and the field parameter $\lambda$. At each step we move
a little bit down the gradient $-\frac{\overline{\delta} {\cal
H}}{\overline \delta \psi}$ and $-\frac{\overline{\delta} {\cal
H}}{\overline \delta \lambda}$. The end product of the iteration is
the self-consistent minimization of the energy. The time dependence
is just a label for different configurations. In practice we chose a
step $\triangle \tau$ and iterate the equations
\begin{equation}
\psi(\tau+\triangle \tau)\approx \psi(\tau) -\triangle
\tau\frac{\overline{\delta} {\cal H}}{\overline \delta \psi(\tau)}~,
\end {equation}
\begin{equation}
\lambda(\tau+\triangle \tau)\approx \lambda(\tau) -\triangle
\tau\frac{\overline{\delta} {\cal H}}{\overline \delta \lambda(\tau)}
\end {equation}
by normalizing $\psi$ and $\lambda$ to the total number of polaritons $N_P$ at
each iteration. The time step $\triangle \tau$ controls the rate of
convergence. The system is described  using 100 points and periodic boundary
conditions and one parameter for the cavity photon field. As a test we compared
our numerical calculations with some analytical limit cases. For the trial
initial  $\psi$ we used a random values on each site and also a form
corresponding to an analytical limit that will be discussed in the next
section. The number of iteration depends on the convergence rate and the choice
of the initial trial function. Typically, we used $10^5-10^6$ iterations.

\section{Results}
\label{sec:chem}
\subsection{Low polariton density}
\begin{figure}
\centering\includegraphics[width=7cm]{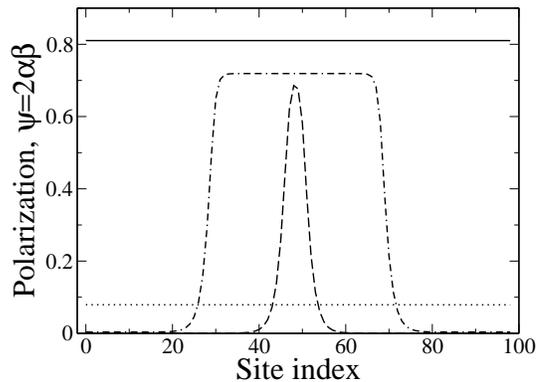} \caption{Polarization
$\psi$ for $\delta/t_0=0.01$, $g/t_0=0.01$, $\chi/t_0=6$, $\rho_
{P}=3\times10^{-3}$ (dotted line), $\rho_{P}=6\times10^{-3}$ (dashed
line), $\rho_{P}=6\times10^{-2}$ (dot-dashed line), and
$\rho_{P}=10$ (solid line).} \label{polar}
\end{figure}
The results of the numerical solution for the function $\psi$ are
shown in Fig.~\ref{polar} for the case $\chi=6t_0$. At very low
excitation densities $\rho_{P}=\frac{N_{P}}{N}$, where $N$ is the
number of sites, polaritons behave as simple bosons. In that limit
the self-trapping effect is absent since the nonlinear attractive
potential is proportional to the local polariton density and can be
neglected in this low density limit. Mathematically, this limit can
be described by approximating
$\frac{\psi}{\sqrt{1-\psi^2}}\approx\psi$  in the last term of
Eq.~(\ref{final}). The cubic term $\psi^3$ is not strong enough to
give rise to the self-trapping effect in this limit.

In the absence of the self-attractive term the distribution of the optical
polarization is homogeneous.  In this low excitation limit the chemical
potential $\mu$ is simply given by
\begin{equation}
\mu=\frac{1}{2}(\omega^\prime_X+\omega_c-\sqrt{\delta^2+g^{\prime2}})~,
\label{chemp}
\end {equation}
where $\omega^\prime_X=\omega_X-2t_0$ (energy at the bottom of the excitonic
band), $\delta=\omega^\prime_X-\omega_c$ is the optical detuning, and
$g^\prime=g\sqrt{N}$. Notice that in this limit the chemical potential does not
depend on the coefficient of the cubic term in Eq.~(\ref{final}). Moreover in
this limit the  chemical potential corresponds to the energy of the lowest
polariton at $k=0$.

\subsection{Intermediate density: self-trapping}
\begin{figure}
\centering\includegraphics[width=7cm]{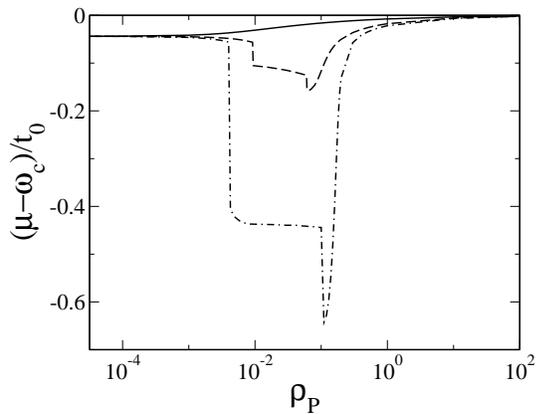} \caption{Polariton
chemical potential as a function of the polariton density for
$\delta/t_0=0.01$, $g/t_0=0.01$, $\chi/t_0=0$ (solid line),
$\chi/t_0=4.5$ (dashed line), and $\chi/t_0=6$ (dot-dashed line).}
\label{chem}
\end{figure}

\begin{figure}[t]
\centering
\includegraphics[width=7cm]{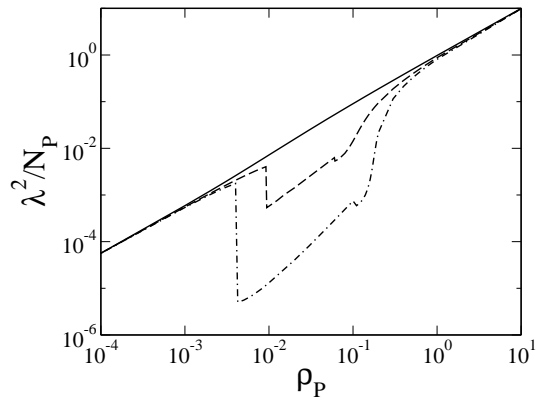}
\caption{Photon density as a function of polariton density for
$\chi/t_0=0$ (solid line), $\chi/t_0=4.5$ (dashed line), and
$\chi/t_0=6$ (dot-dashed line).}\label{lambda}
\end{figure}

By increasing the number of polaritons  we observe a critical
polariton number at which a symmetry-breaking transition occurs. A
similar symmetry-breaking transition was found in the case of BEC
with an attractive nonlinear interaction.~\cite{carr00} Our
numerical calculations for the dependence of $\mu$ on the polariton
density $\rho_{P}=\frac{N_{P}}{N}$ are shown in Fig.~\ref{chem} for
several values of the cubic term $\chi$.
When the exciton number increases following the polariton number, a
self-trapping occurs, caused by the polariton-lattice coupling
(dashed line in Fig.~\ref{polar}). There is a drop in the chemical
potential at the point of symmetry-breaking (Fig.~\ref{chem}) due to
the self-trapping. The breaking of the spatial homogeneity gives a
negative contribution to the energy of the system. Also, this makes
exciton-like states more energetically favorable than photon-like
states, which results in a sharp decrease of the photonic component
for nonzero $\chi$, as shown in Fig.~\ref{lambda}. The self-trapping
effect brings a discontinuity both in the chemical potential $\mu$,
and in the photon (exciton) density as a function of the polariton
density. Since the effective attractive potential depends on both
$|\psi|^2$ and $\chi$, the point of symmetry breaking for higher
$\chi$ corresponds to smaller values of the polariton density
$\rho_{P}=\frac{N_{P}}{N}$.

Starting from the small excitation limit, we can expand
$\frac{\psi}{\sqrt{1-\psi^2}}\approx\psi+\frac{1}{2}\psi^3$. The
second term introduces an effective repulsion related to the
intrinsic fermionic nature of the excitons and decreases the cubic
term $\chi$ by $\frac{\omega_X-\mu}{2}$. Some analytical forms for
the solution of Eq.~(\ref{eq:fina}) with fixed exciton number can be
written in terms of Elliptical functions. In particular, in the case
when the photon number is much smaller than the exciton number, we
can neglect the fourth term in Eq.~(\ref{eq:fina}), since as seen
from Eq.~(\ref{eq:finb}), $\lambda$ is inversely proportional to
$\omega_c -\mu$. In this limit the solution can be written in the
form
\begin{equation}
\psi=A\mathrm{sech}[Y(\nu-\nu_0)]~,
\end {equation}
where
\begin{equation}
Y=\frac{(\chi-\frac{\omega_X-\mu}{2})N_{P}}{4t_0}~,
\end {equation}
\begin{equation}
 A=\sqrt{\frac{Y N_{P}}{2}}~,
\end {equation}
 and the chemical potential is determined by the equation
\begin{equation}
 \mu=\omega_X-2t_0-\frac{(\chi-\frac{\omega_X-\mu}{2})^2N^2_{P}}{16t_0}~.
\end {equation}

\subsection{Saturation}
\begin{figure}
\centering
\includegraphics[width=7cm]{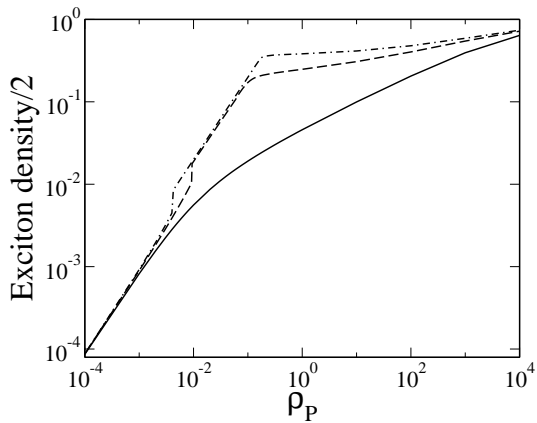}
\caption{Half of exciton density as a function of the polariton
density for $\chi/t_0=0$ (solid line), $\chi/t_0=4.5$ (dashed line),
and $\chi/t_0=6$ (dot-dashed line).}\label{exciton}
\end{figure}

\begin{figure}
\centering
\includegraphics[width=7cm]{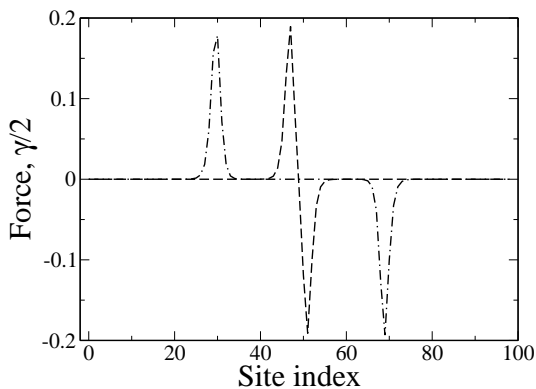}
\caption{Force as a function of the site index for $\chi/t_0=6$,
$\rho_{ex}=6\times10^{-3}$ (dashed line), $\rho_{ex}=6\times10^{-2}$
(dot-dashed line).} \label{force}
\end{figure}

By further increasing the polariton density, the internal fermionic
structure of the excitons gives rise to a hard-core repulsion term,
which leads to the saturation of the exciton states, making the
polarization distribution broader and flatter at the top (dot-dashed
line in Fig.~\ref{polar}). When the saturation spreads over the
whole chain length, the polarization distribution (solid line in
Fig.~\ref{polar}) becomes homogeneous again. We expect a
discontinuity of $\mu$ again at this point as a consequence of the
disappearance of the gradient of the polarization function (kinetic
term), which gives a positive contribution to the energy. With the
further increase of the exciton density the hopping effect is
reduced due to the blocking. Fig.~\ref{exciton} shows half of the
exciton density $\frac{\int|\beta|^2d\nu}{N}$ as a function of
$\rho_{P}$, which  reaches 1 in the limit of large excitation
density regardless the value of $\chi$, corresponding to half
filling of the exciton band. The half filling maximizes the
polarization and hence minimizes the dipole interaction energy
between the excitons and the cavity photons. In this saturation
regime the polaritons become photon-like, since an added excitation
mainly contributes to the cavity mode, and thus the chemical
potential approaches $\omega_c$ (see also Fig.~\ref{chem}). A
gradient in the density of polarization produces a force on the ions
according to Eq.~(\ref{eq:eub}). The force is stronger at the edges
of the saturation region of the $|\psi|^2$ distribution as seen in
Fig.~\ref{force}, and is positive to the left from the center of the
symmetry-breaking point and negative to the right. This reduces of
the total length of the chain due to the interaction with the
electromagnetic cavity mode.


\section{Homogeneous deformation and role of damping}
\label{sec:bistab} The lattice deformation in the presence of
optical excitations can be understood by analyzing the total energy
of the system in the assumption that the lattice deformation is
homogeneous. As seen in the previous section, this assumption is
justified for a systems in the saturation regime or for a polariton
density below the critical value for self-trapping. The total energy
of the lattice is modified by the presence of polaritons. The
homogeneous deformation allows us to obtain an analytical expression
for the total energy of the system, and to analyze the effect of the
finite linewidth of polaritons. We will include both the spontaneous
emission of the excitons and the finite Q-factor of the cavity mode.
Instead of using the variational approach of Sec.~\ref{sec:energy},
we will solve directly the equations of motion for the excitons and
cavity photons.

We start by using the Fourier transform
\begin{equation}
B_n=\frac{1}{\sqrt{N}}\sum\limits_k e^{ikna}b_k~,
\end{equation}
where $N$ is the number of lattice sites, $k$ is the wave vector,
and $a$ is the lattice separation, to define the operator $b_k$ as
the annihilation operator of an exciton with a wave vector $k$. The
Hamiltonian in Eq.~ (\ref{eq:htot}) can then be rewritten as
\begin{eqnarray}
H=\sum\limits_n\left[\frac{p_n^2}{2M}+ \frac{C}{2}\left(u_{n+1}-u_n\right)^2
\right]+ \nonumber \\ + \sum\limits_k \omega_k b^\dagger_k b_{k}  +
\sum\limits_{k,k'} f(k,k') b^\dagger_k b_{k'} \nonumber \\+
  \omega_c c^\dagger c + \frac{g'}{2} c^\dagger b_0+\frac{g'}{2} c
  b^\dagger_0~,
\label{eq:hamk}
\end{eqnarray}
where $\omega_k=\omega_X-2t_0\cos(ka)$ and
\begin{eqnarray}
f(k,k')=\frac{\gamma}{N}\sum\limits_n\left(u_{n+1}-u_n \right)
\times \nonumber \\ \times
\left[e^{-ika}e^{ina(k-k')}+e^{ik'a}e^{ina(k'-k)} \right]~.
\end{eqnarray}
We consider a finite-length chain with periodic boundary conditions. The
equilibrium lattice displacement is homogeneous and will be indicated by
$u=u_{n+1}-u_n$. Notice that in this limit $f(k,k')$ is diagonal
\begin{equation}
f(k,k')=2 \gamma \cos(ka) u \delta_{k,k'}~. \label{fkk}
\end{equation}
Eqs.~(\ref{fkk}) and~ (\ref{eq:hamk}) show that in this homogeneous
case there is no mixing of polariton with different $k$ vector.
Since the cavity couples to the $k=0$ exciton mode only, the
polariton modes at $k \ne 0$ are completely decoupled and do not
enter in the dynamics.  Taking into account Eq. (\ref{fkk}), the
excitonic part in $H$ reduces to $\omega_0(u) b_0^\dagger b_0$,
where $\omega_0(u)=\omega_X-2 t_0+2\gamma u$ is the energy of the
exciton at $k=0$ renormalized by the lattice deformation potential.
We also add a term that describes the pumping of the cavity mode by
an external field and represent the full system by
\begin{eqnarray}
H=\sum\limits_n\left[\frac{p_n^2}{2M}+ \frac{C u^2}{2}
\right]+ \nonumber \\
+\omega_0 (u)b_0^\dagger b_0+\frac{g'}{2}(c^\dagger b_0+b_0^\dagger
c)+\omega_cc^\dagger c+ \nonumber \\
\frac{\kappa E_0}{2}(c e^{i\omega_c t}+c^\dagger e^{-i\omega_c t} )~,
\label{tham}
\end{eqnarray}
where $E_0$ represents the electric field of the external pump and $\kappa$ is
the coupling between the external pump and the cavity mode. The operators $B_n$
are fermionic on the same site, but commute for different sites. Therefore, the
commutation relation for $b_0$ reads
\begin{equation}
[b_0,b_0^\dagger]=1-\frac{2\sum\limits_nB_n^\dagger B_n}{N}~.
\end{equation}
Since we will assume here that there are no excitations of phonon
modes at $k \ne 0$, we have that $\sum\limits_N B_n^\dagger
B_n=\sum\limits_k b_k^\dagger b_k \sim b_0^\dagger b_0$. The
equations for the expectation values of the polarization $\langle
b_0 \rangle=p$, exciton density $\langle b^\dagger_0 b_0
\rangle=n_X$ and cavity photon operator $\langle c \rangle=\lambda$
can be obtained using the standard factorization scheme~\cite{axt94}
and read
\begin{equation}
\dot \lambda=-i\left[\omega_c \lambda+\frac{g'}{2}p+\frac{\kappa
E_0}{2}e^{-i\omega_c t} \right]-\alpha_c \lambda~,
\end{equation}
\begin{equation}
\dot p =-i\left[\omega_0 (1-\frac{2 n_X}{N})p+ \frac{g'}{2} \lambda
(1-\frac{2 n_X}{N})\right]-\Gamma p~, \label{eq:beta}
\end{equation}
\begin{equation}
\dot{n_X}=-i\frac{g'}{2} (1- \frac{2 n_X}{N})(p^*\lambda-p
\lambda^*)-2\Gamma n~, \label{eq:n}
\end{equation}
where we have introduced the spontaneous emission rate of the
excitons $\Gamma$ and the damping of the cavity mode $\alpha_c$, due
to the finite $Q$-factor of the cavity. The equation for $ \lambda$
can be explicitly integrated and gives
\begin{equation} i \lambda=\frac{g'}{2}\frac{p}{\alpha_c}+ \frac{\kappa E_0}{2\alpha_c}e^{-i\omega_c
t}~.
\end{equation}
This expression can be used in Eqs.~(\ref{eq:beta}) and
~(\ref{eq:n}) which give
 for the steady state
 \begin{equation}
\tilde p =\frac{-i \frac{g'}{2}\frac{\kappa E_0}{2\alpha_c}
(1-\frac{2 n_X }{N})}{ \omega_c - \omega_0 (1-\frac{2 n_X
}{N})+i\Gamma +i\frac{g'^2}{4\alpha_c} (1-\frac{2 n_X)}{N})}~,
\end{equation}
where we have defined $p=\tilde p e^{-i\omega_c t}$.

\begin{figure} \centering \includegraphics[height=7.cm,angle=0]{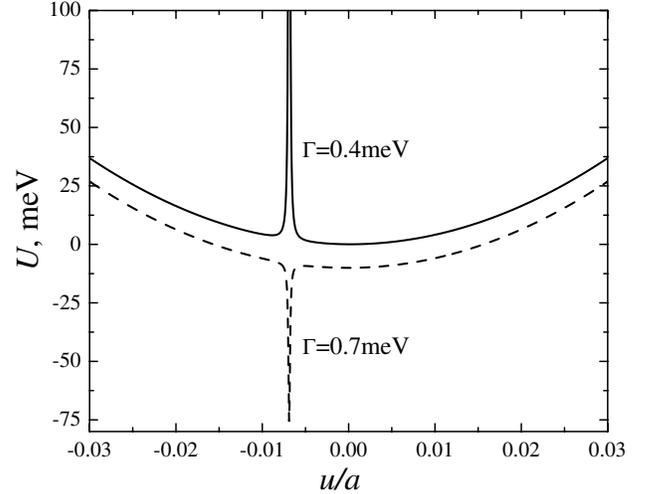}
\caption{Effective lattice potential as a function of displacement.
The figure has been obtained for the following set of parameters:
$E_g=\omega_X-2t_0=2.282$ eV, $D=2 \gamma a=6.1$ eV, $S=2.5\cdot
10^3$m/s, $a=1.5$ nm, $M= 2.1 10^{-21}$ g, $\omega_0=2.282$ eV, and
$C=S^2 M/a^2=82 eV/a^2$. The excitation corresponds to
$\rho_{ex}\sim 0.1$. $\alpha_c=g'= 1$ meV, and $\omega_c=2.242$ eV.
$\Gamma=400$ $\mu$eV and  $\Gamma=700$ $\mu$eV for the upper and
lower curve, respectively. The curves are displaced for clarity.
\label{bistab}}
\end{figure}

The $u$-dependent total potential energy $U(u)$ of the system can be
obtained by substituting the solution for $p$, $\lambda$, and $n_X$
which depend on $u$, into the initial Hamiltonian(\ref{tham}). At
weak excitation we obtain
\begin{eqnarray}
U(u)\sim\frac{NC}{2}u^2+ \omega_0(u) n_X(u)+ \nonumber \\
+\omega_c\left[\left(\frac{g'}{2 \alpha}\right)^2-\frac{2
\Gamma}{\alpha_c}\right]n_X(u) +\omega_c I~ \label{eq:pulling}
\end{eqnarray}
where $I=\left(\frac{\kappa E_0}{2 \alpha_c}\right)^2$ and
\begin{equation}
n_X(u)\sim |p|^2= \frac{I (g'/2)^2}{(\delta+2\gamma u)^2+\Gamma^2}~.
\label{eq:res}
\end{equation}
Notice that $n_X$ has a resonant behavior as a function of $u$,
which is a consequence of the deformation potential shift in the
excitonic energy.

In the absence of light the potential energy $U$ depends
quadratically on $u$ with a minimum at $u=0$. When the laser is
switched on, this dependence changes due to the presence of optical
excitations in the system.  Due to the resonant behavior in Eq.
~(\ref{eq:res}), it is energetically favorable to have $u \ne 0$ if
this reduces the total energy of the system. Figure~ \ref{bistab}
shows the potential energy for some values of parameters. The
parameters used in these calculations are typical of polydiacetylene
chains. We observe that the potential $U$ has a parabolic dependence
with a superposed Lorentian contribution due to the optical
excitation. This Lorentian contribution can be either positive or
negative depending on the relative strength of the exciton-cavity
coupling, spontaneous emission rate and cavity damping. The actual
behavior can be explicitly calculated using Eq.~(\ref{eq:pulling}).
In the figure we consider two particular sets of parameters
providing a minimum of the energy that corresponds to contraction
and expansion of the lattice.

\section{Conclusion}
\label{sec:conclusions} In conclusions, we have investigated the
optically-induced lattice strain in a single polymer chain in a
cavity. We have extended the excitonic SSH model, describing the
exciton propagation in the chain, to include the effect of the
cavity electromagnetic field. Using a polariton picture, we have
obtained a system of integro-differential nonlinear equations for
the spatial distribution of the optical excitations in the chain. We
find solutions describing a self-trapping of the polaritons which
saturate when the excitation density is increased. The chemical
potential of the polaritons shows a discontinuity at a threshold
value of the polariton density. This critical density corresponds to
the onset of the self-trapping. The self-trapping causes a sharp
increase in the excitonic component and decrease in the photonic
component of the polariton wavefunction. We have also considered the
role of the finite radiative recombination rate of the excitons and
the finite Q-factor of the cavity. These  can be studied in a direct
way in the case of a homogeneous strain field. We have found that
both a contraction and expansion of the lattice are possible,
depending of the relative strength of the exciton-cavity coupling,
radiative recombination, and cavity Q-factor.

\section*{Acknowledgments}
This research was supported by the National Science Foundation, Grant NSF
DMR-0312491 and DMR-0605801.

\end{document}